\documentclass[twocolumn,showpacs,preprintnumbers,amsmath,amssymb]{revtex4}
\usepackage{graphics,epsfig,subfigure}
\usepackage{color}

\begin{document}
\renewcommand{\baselinestretch}{1.3}
\newcommand\beq{\begin{equation}}
\newcommand\eeq{\end{equation}}
\newcommand\beqn{\begin{eqnarray}}
\newcommand\eeqn{\end{eqnarray}}
\newcommand\nn{\nonumber}
\newcommand\fc{\frac}
\newcommand\lt{\left}
\newcommand\rt{\right}
\newcommand\pt{\partial}
%\baselineskip15pt

%\title{A warped description of large extra dimension}
%\title{The hierarchy problem and new extra dimension at a few TeV$^{-1}$}
%{Describing the ADD model in a warped geometry}
\title{The Hierarchy Problem and New Warped Extra Dimension}

%  New solution to gauge hierarchy problem
%  New solution to gauge hierarchy problem
% Gauge hierarchy problem and new extra dimension model
% Gauge hierarchy problem and new extra dimension mechanism
% The hierarchy problem and new extra dimension at a TeV$^{-1}$

\author{Bin Guo$^{1}$\footnote{These authors contributed equally to this work.} }%\email{guob12@lzu.edu.cn}
\author{Yu-Xiao Liu$^{1,2}$\footnote{Corresponding author. Email: {liuyx@lzu.edu.cn}}
\footnote{These authors contributed equally to this work.}}
\author{Ke Yang$^{1,3}$}%\email{yangke09@lzu.edu.cn}
\author{Shao-Wen Wei$^{1,2}$}
\affiliation{
$^{1}$Institute of Theoretical Physics $\&$ Research Center of Gravitation, Lanzhou University, Lanzhou 730000, China\\
$^{2}$Key Laboratory for Magnetism and Magnetic of the Ministry of Education, Lanzhou University, Lanzhou 730000, China\\
$^{3}$School of Physical Science and Technology, Southwest University, Chongqing 400715, China}

\begin{abstract}
In this paper, we propose a new mechanism with warped extra dimension to solve the hierarchy problem,  which is parallel to the Randall-Sundrum (RS) brane scenario.
Different from the RS scenario, the fundamental scale is TeV scale and the
four-dimensional Planck scale is generated from the exponential
warped extra dimension at size of a few TeV$^{-1}$.
{ The experimental consequences of this scenario are very different from that of the RS scenario.
In the explicit realization in the nonlocal gravity theory, there is a tower of spin-2 excitations with mass gap $10^{-4}\text{eV}$ and they are coupled with the gravitational scale to the standard model particles. We further discuss the possible generalizations in other modified gravity theories. The experimental consequences are similar to $(4+N)$-dimensional large extra dimension but $N$ can be a non-integer, which satisfies the experimental constraints more easily than the integer large extra dimension model. }

\end{abstract}


% 04.20.Jb Exact solutions
% 04.50.-h Higher-dimensional gravity and other theories of gravity
% 04.50.Kd Modified theories of gravity
% 04.60.Rt Topologically massive gravity
%           (see also 11.15.Wx Topologically massive gauge theories,
%           and 11.15.Yc Chern-Simons gauge theory)
%04.20.-q Classical general relativity (see also 02.40.-k Geometry, differential geometry, and topology)
%14.70.Kv Gravitons (see also 04.60.-m Quantum gravity)

% 04.60.-m Quantum gravity (see also 11.25.-w Strings and branes; 11.15.Wx Topologically massive gauge theories, and 11.15.Yc Chern-Simons gauge theory)

% 04.50.+h Gravity in more than four dimensions, Kaluza-Klein theory,
           % unified field theories, alternative theories of gravity
           %(see also 11.25.M Compactification and four-dimensional models), dilaton gravity
% 11.27.+d Extended classical solutions; cosmic strings,
           %domain walls, texture (see 98.80.C in cosmology)

\maketitle

\section{Introduction}\label{introduction}

It is known that the gauge hierarchy problem, the large difference between the electroweak scale $M_{\text{EW}}\sim 1$TeV and the Planck scale $M_{\text{Pl}}\sim 10^{16}$TeV, is
a longstanding problem in particle physics.
The idea of extra dimensions opens a new way to solve this problem. One of the famous models is the Arkani-Hamed-Dimopoulos-Dvali (ADD) model \cite{ArkaniHamed:1998rs, Antoniadis:1998ig}, also called the model of large extra dimensions.

In this model, the fundamental scale is $M_{*}\sim 1$TeV. The standard model particles are assumed to be confined on the brane and the extra dimensions are flat, so the electroweak scale is the same as the fundamental scale: $M_{\text{EW}}\sim M_{*}$.
The hierarchy between the effective Planck scale $M_{\text{Pl}}$ and the fundamental scale $M_{*}$ is generated by the large volume of the extra dimensions \cite{ArkaniHamed:1998rs}:
\begin{eqnarray}
M^{2}_{\text{Pl}}\sim M^{N+2}_{*}R^{N},
\end{eqnarray}
where $R$ and $N$ are the size and number of the extra dimensions, respectively.

However, the gauge hierarchy between the Planck and electroweak scales has not been solved ultimately because it has been transferred into a remained hierarchy between the fundamental length $M_{*}^{-1}\sim 2\times10^{-16}$mm and the size of the extra dimensions $R \sim 10^{{32}/{N}} M_{*}^{-1}$ (about 2mm for the case of two extra dimensions).
Furthermore, the brane tension is neglected in this model.

A good inspiration to solve the remained hierarchy problem in the ADD model comes from the Randall-Sundrum 1 (RS1) model \cite{Randall:1999ee} in five dimensions with the warped geometry given by
\begin{eqnarray}
ds^{2}&=&e^{-2k|y|}\eta_{\mu\nu}dx^{\mu}dx^{\nu}+dy^{2}. \label{RS1}  %&=&\frac{1}{(1+k|z|)^2}\big(\eta_{\mu\nu}dx^{\mu}dx^{\nu}+dz^{2}\big), \label{RS1b}
\end{eqnarray}
Here $k$ is the AdS curvature scale, $y\in [-y_{b},y_{b}]$ is the physical coordinates of the fifth dimension.

\begin{figure}[htbp]
\subfigure[~RS1 mechanism]{\label{picRS1}
\includegraphics[scale=0.45]{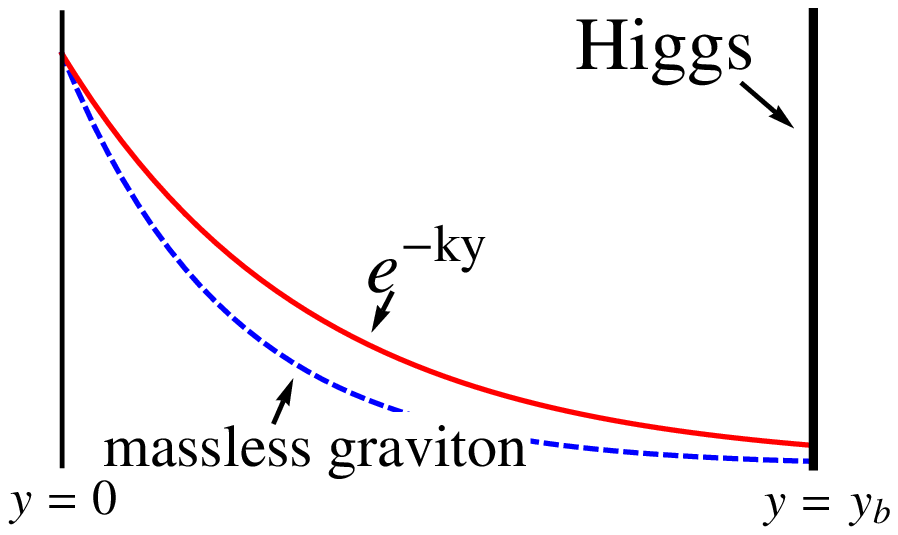}}
\subfigure[~New mechanism]{\label{picnewmechanism}
\includegraphics[scale=0.45]{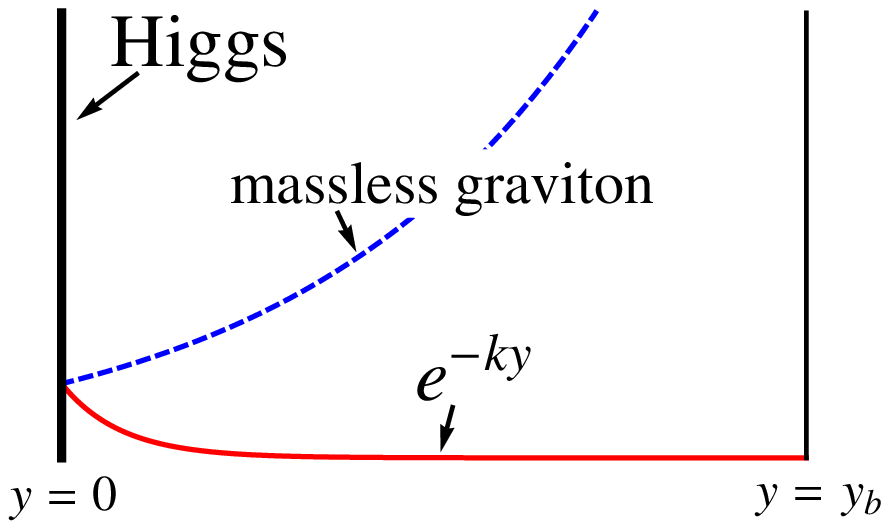}}
\caption{The pictures of the RS1 and the new mechanism.}\label{picmasslessgraviton}
\end{figure}

The fundamental scale is $M_{*} \sim k \sim 10^{16}$TeV $=M_{\text{Pl}}$ in the RS1 model.
Because the massless graviton is localized near $y=0$ as illustrated in figure \ref{picRS1} and it has not been diluted exponentially by the size of the extra dimension, the effective Planck scale is the same as the fundamental scale.
On the other hand, in background (\ref{RS1}) the mass parameters of fields confined at $y=y_0$ has a redshift $e^{-ky_{0}}$, so the standard model particles should be confined at $y=y_{b}$ with $y_{b}\sim 37 {k}^{-1} \sim 37 M_{*}^{-1}$ to recover the electroweak scale.
The gauge hierarchy problem is solved by a small physical length of the fifth dimension $y_{b}\sim 37 M_{*}^{-1}$, for which there is no remained hierarchy problem.

After the success of the RS1 model, many researchers constructed the extended RS1 models in modified gravities. The main idea of these extended models is the same as the RS1 model's, i.e., the massless graviton is localized near $y=0$ and the standard model particles are confined at $y=y_{b}$. In this { paper} we will show that another mechanism can also be used to solve the gauge hierarchy problem. In this new mechanism, the massless graviton is localized near $y=y_{b}$ and the standard model particles are confined at $y=0$ as illustrated in figure \ref{picnewmechanism}. This will lead to new extra dimension at a few TeV$^{-1}$ and new phenomena. We first make an explicit realization of this mechanism in nonlocal gravity. Then, we realize this mechanism in a more general class of modified gravity theories.

\section{The model}\label{the model}

Different from the usual way to modify Einstein gravity by local generalization, Deser and Woodard developed a modified gravity theory by adding a nonlocal term \cite{Deser:2007jk, Woodard:2014iga}.
This nonlocal gravity theory was proposed to explain the cosmic acceleration without the cosmological constant \cite{Deffayet:2009ca,Koivisto:2008xfa,Nojiri:2010pw,Elizalde:2012ja,Elizalde:2013dlt}.
It was found that this theory possesses the same gravitational degrees of freedom and initial value constraints as general relativity, and has no ghost graviton mode \cite{Deser:2013uya}.
We start with the following action of the nonlocal gravity theory proposed in Ref.~\cite{Deser:2007jk} in $(d+1)$-dimensional spacetime:
\begin{eqnarray}
S= \frac{M^{d-1}_{*}}{2}
  \int d^{d+1}x \sqrt{-g}
  \, \mathcal{R} \Big(1+f(\Box^{-1}\mathcal{R})\Big)
          +S_{\texttt{b}},\label{nonlocal form}
\end{eqnarray}
where $M_{*}$ is the fundamental $(d+1)$-dimensional Planck mass, $\mathcal{R}$ is the curvature scalar, $\Box^{-1}$ is the inverse of the d'Alembertian operator, and $S_{\texttt{b}}$ is the action describing the brane configuration. Note that in order to define this theory, we should specify what definition of $\Box^{-1}$ we use.
A convenient way to solve the nonlocal theory is changing to the local form by adding two auxiliary fields $\eta$ and $\xi$ \cite{Nojiri:2007uq,Elizalde:2012ja,Elizalde:2013dlt}:
\begin{eqnarray}
S = \frac{M^{d-1}_{*}}{2}\int d^{d+1}x \sqrt{-g}\left({\psi}\mathcal{R}+\xi\Box\eta\right)
   +S_{\texttt{b}},   \label{local form}
\end{eqnarray}
where we have defined $\psi\equiv1+f(\eta)-\xi$.
To ensure the stability, we should require $\psi>0$.
By varying the action (\ref{local form}) with respect to $\xi$ and $\eta$ respectively, one {gets}
\begin{eqnarray}
\Box \eta = \mathcal{R},  ~~~
\Box \xi = -\mathcal{R} f_{\eta}, \label{abs eta xi}
\end{eqnarray}
where $f_{\eta}\equiv \frac{d f}{d \eta}$.
Now the choice of the definition of $\Box^{-1}$ becomes the choice of the solution of $\Box \eta = \mathcal{R}$. Once we have chosen a specific solution of $\eta$, the original nonlocal theory is defined. It means that different solutions of $\eta$ correspond to different nonlocal theories, rather than different solutions of one nonlocal theory.

The modified Einstein equations are given by
\begin{eqnarray}
&& \psi \mathcal{R}_{AB}
  - \frac{1}{2}\left({\psi}\mathcal{R}-\partial_{C}\xi\partial^{C}\eta\right) g_{AB}
   \nonumber\\
&+& (g_{AB}\Box-\nabla_{A}\nabla_{B})\big(\psi-1\big)
 - \frac{1}{2} \big(\partial_{A}\xi\partial_{B}\eta
                     +\partial_{A}\eta\partial_{B}\xi \big)      \nonumber\\
&=& M^{1-d}_{*} T^{(\texttt{b})}_{AB}, \label{background equation}
\end{eqnarray}
where $A, B, C\cdots$ denote the bulk indices $0,1,2,\cdots,d$, and $T^{(\texttt{b})}_{AB}$ represents the energy-momentum tensor of the brane.

The metric for the flat brane world model is \cite{Randall:1999ee}
\begin{eqnarray}
ds^{2}=a^{2}(y)\eta_{\mu\nu}dx^{\mu}dx^{\nu}+dy^{2}, \label{background metric}
\end{eqnarray}
where the brane coordinate indices $\mu,\nu=0,1,\cdots,d-1$.
We consider an $S_{1}/Z_{2}$ orbifold extra dimension as in Ref. \cite{Randall:1999ee}, so the physical extra dimensional coordinate $y\in [-y_{b},y_{b}]$. The action describing one brane with tension $\sigma_{-}$ located at $y=0$ and another brane with tension $\sigma_{+}$ located at $y=y_{b}$ is given by
\begin{eqnarray}
S_{\texttt{b}}=-\int d^{d}x
   \Big[ \sqrt{-q({y=0})} \;\sigma_{-}
        +\sqrt{-q(y=y_{b})}\;\sigma_{+}
    \Big],
\end{eqnarray}
where $q_{\mu\nu}(y=0,y_b)$ are the induced metrics at $y=0,y_{b}$.

Then the $\mu\nu$ and $dd$ components of the modified Einstein equations (\ref{background equation}) and the equations (\ref{abs eta xi}) for $\eta$ and $\xi$ are given by
\begin{eqnarray}
    \psi'' \!+\!(d\!-\!1)H \psi'
    \!+\!\big[(d\!-\!1)H'\!+\!\frac{1}{2}d(d\!-\!1)H^{2}\big]\psi   \!\! \nonumber \\
       +\frac{1}{2}\xi'\eta'
  +{M^{1-d}_{*}} \big[\sigma_{-}\delta(y)+\sigma_{+}\delta(y-y_{b})\big]
  \!\!&=&\!\!0,\label{Einstein uv}  \\
   \frac{1}{2}\xi'\eta' - d H \psi' - \frac{1}{2}d(d-1)H^{2}\psi  \!\!&=&\!\! 0, ~~\label{Einstein DD}
\end{eqnarray}
and
\begin{eqnarray}
\eta''+d H \eta' \!&=&\! -2 d H'-d(d+1)H^{2},     \label{eta}\\
\xi''+d H \xi' \!&=&\! \big[2d H'+d(d+1)H^{2}\big]f_{\eta},  \label{xi}
\end{eqnarray}
respectively, where $H\equiv{a'}/{a}$ and the prime denotes the derivative with respect to
the coordinate $y$. Subtracting (\ref{Einstein DD}) from (\ref{Einstein uv}) we get
\begin{eqnarray}
  &&\psi''+(2d-1)H \psi'+\big[(d-1)H'+d (d-1) H^{2}\big]\psi \nonumber\\
 &=& -{M^{1-d}_{*}} \big[\sigma_{-}\delta(y)+\sigma_{+}\delta(y-y_{b})\big], \label{eq d}
\end{eqnarray}
which can be rewritten as
\begin{eqnarray}
  &&\Big(\partial_{y}+d H\Big)\Big(\partial_{y}+(d-1) H\Big)\psi \nonumber\\
 &=& -{M^{1-d}_{*}} \big[\sigma_{-}\delta(y)+\sigma_{+}\delta(y-y_{b})\big]. \label{integrable eq}
\end{eqnarray}
We consider the warp extra dimension with $a(y)=e^{-k|y|}$. Then $\psi$ can be { obtained} as
\begin{eqnarray}
\psi=c_{1} e^{d k |y|}+c_{2} e^{(d-1)k|y|}.
\end{eqnarray}
The condition $\psi>0$ becomes $c_{2}>-c_{1}e^{k|y|}$.
After integrating Eq.~(\ref{integrable eq}) at $y=0$ and $y=y_{b}$ respectively, we can obtain the relations
\begin{eqnarray}
   c_{1}=-{\sigma_{-}}/{(2 k M^{d-1}_{*})},~~\sigma_{+}=-\sigma_{-}e^{d k y_b}.
   %\sigma_{+}=-\sigma_{-}e^{-d k y_b}.
\end{eqnarray}
We assume that the brane located at $y=0$ has negative tension, i.e., $\sigma_{-}<0$.
Then the brane located at $y=y_{b}$ is a positive tension brane.
The relation between the $\sigma_{+}$ and $\sigma_{-}$ is different from the case in the RS1 model, where $\sigma_{+}=-\sigma_{-}$.

The solutions of the auxiliary fields $\eta$ and $\xi$ are complicated.
We will use Eq. (\ref{eta}) for $\eta$ as an example.
We denote the region $y\in[n y_{b},(n+1)y_{b}]$ as the $n$-th section in the region $y\in[0,+\infty)$ and denote the solution of the field $\eta(y)$ in this section as $\eta_{n}(y)$.
Then the solution of Eq. (\ref{eta}) is given by
\begin{eqnarray}
\eta_{n}(y)=(-1)^n (d+1) k y + \lambda_{n}e^{(-1)^n dk(y-n y_{b})}+\rho_{n},~
\end{eqnarray}
where the coefficients $\lambda_{n}$ are determined by the recursion relation
$
\lambda_1={2(d-1)}/{d}-e^{dky_{b}} \lambda_0$,
$\lambda_{n+2}=\big(1-e^{-(-1)^n dky_{b}}\big){2(d-1)}/{d}+\lambda_{n}.
$
The coefficients $\rho_{n}$ can be determined by the continuity condition of $\eta(y)$.
The exact solutions of the auxiliary fields $\eta$ and $\xi$ are unnecessary since they do not influence the spectrum of the massive KK gravitons and the couplings of these gravitons to matter.

\section{Physical implications}\label{physical implication}

{ Since} the metric (\ref{background metric}) has the $d$-dimensional Poincare symmetry, we can consider the transverse and traceless (TT) part of the perturbation separately, which corresponds to the spin-2 graviton in $d$ dimensions.
The tensor perturbation is parameterized as
\begin{eqnarray}
ds^{2} =a^{2}(\eta_{\mu\nu}+h_{\mu\nu})dx^{\mu}dx^{\nu}+dy^{2},
\end{eqnarray}
where the perturbation $h_{\mu\nu}(x,y)$ satisfies the TT conditions
\begin{eqnarray}
\partial^{\mu}h_{\mu\nu}=0=\eta^{\mu\nu}h_{\mu\nu}.
\end{eqnarray}
The $\mu\nu$ component of the linearized perturbation equations reads
\begin{eqnarray}
a^{2}h''_{\mu\nu}+\Big(d a a'+a^{2}\frac{\psi'}{\psi}\Big)h'_{\mu\nu}+\Box^{(d)}h_{\mu\nu}=0,
\label{linearEq1}
\end{eqnarray}
where $\Box^{(d)}$ denotes the $d$-dimensional d'Alembertian operator.
After changing to the conformal coordinate $z$ with $a dz=dy$
and defining $A\equiv a^{d-1}\psi$, the linearized equation (\ref{linearEq1}) becomes
\begin{eqnarray}
\partial_{z}\partial_{z}h_{\mu\nu}
 +\frac{\partial_{z}A}{A}\partial_{z}h_{\mu\nu}+\Box^{(d)}h_{\mu\nu}=0.
 \label{linearEq2}
\end{eqnarray}
We decompose the tensor perturbation $h_{\mu\nu}$ as
\begin{eqnarray}
h_{\mu\nu}(x,z)=\varepsilon_{\mu\nu}(x)A^{-\frac{1}{2}}(z)\Psi(z),
\end{eqnarray}
where the four-dimensional part of the graviton KK mode $\varepsilon_{\mu\nu}(x)$ satisfies  { the $d$-dimensional Klein-Gordon equation}
\begin{eqnarray}
\Box^{(d)}\varepsilon_{\mu\nu}(x)=m^{2}\varepsilon_{\mu\nu}(x).
\end{eqnarray}
Then we obtain a Schr\"{o}dinger-like equation for the fifth dimensional part:
\begin{eqnarray}
-\partial_{z}\partial_{z}\Psi+\left(\frac{1}{2}\frac{\partial_{z}\partial_{z}A}{A}
-\frac{1}{4}\frac{(\partial_{z}A)^{2}}{A^{2}}\right)\Psi=m^{2}\Psi. \label{Shrodinger-like eq}
\end{eqnarray}
{ It can be rewritten as the form of the supersymmetric quantum mechanics
\begin{eqnarray}
\mathcal{ K}^\dagger\mathcal{K}\Psi=m^{2}\Psi,
  \end{eqnarray}
where $ \mathcal{ K}^\dagger=\partial_{z}+\frac{1}{2}\frac{\partial_{z}A}{A}$ and $\mathcal{K}=-\partial_{z}+\frac{1}{2}\frac{\partial_{z}A}{A}$. This} implies that the eigenvalues $m^{2}$ are nonnegative and the brane system is stable under the tensor perturbation.

From now on we focus on the case of $d=4$.
Setting $m^{2}=0$ in Eq. (\ref{Shrodinger-like eq}), we can obtain the zero mode
\begin{eqnarray}
  \Psi_{0}(z)\propto {A^{\frac{1}{2}}}
  =\left({\frac{\sigma_{+}}{2 k M^{3}_{*}}(1+k|z|)+c_{2}}
   \right)^{\frac{1}{2}}\sim e^{\frac{1}{2}k y},
\end{eqnarray}
which is localized near
$y=y_{b}$, as expected. { So in order to achieve a weak four-dimensional effective gravity, we should put our universe on the brane at origin, where the massless graviton is diluted exponentially into extra dimension.} The Planck mass is warped up:
\begin{eqnarray}
M_{\text{Pl}}^{2}=M_{*}^{3}\int dz A = M_{*}^{3} z_{b}\left[\frac{\sigma_{+}}{2 k M^{3}_{*}}(2 +k z_{b})+2 c_2 \right].\label{Planck mass}
\end{eqnarray}

On the other hand, as in the RS1 model the mass parameter of a field confined at $y=y_{0}$ has a redshift $e^{-ky_{0}}$ \cite{Randall:1999ee}. So the simplest way to solve the gauge hierarchy problem is to choose the fundamental scale as 1TeV and confine the standard model particles at $y=0$. Thus, the electroweak scale remains the fundamental scale $M_{\text{EW}}\sim1$TeV. In detail, we set $M_{*}\sim k\sim 1$TeV, $\sigma_{+}\sim (1\text{TeV})^{4}$ and the dimensionless parameter $c_2\sim 1$.
Then the effective four-dimensional Planck mass $M_{\text{Pl}}\sim 10^{16}$TeV leads to the condition $k z_{b}\sim 10^{16}$, which means we need only a small physical length $k y_{b}\sim 37$ (or $y_b \sim 37 \text{TeV}^{-1}$) to solve the gauge hierarchy problem. This is a new extra dimension at a few TeV$^{-1}$, comparing to a few $M_{\text{Pl}}^{-1}$ in the RS1 model.

In the following discussion, we check whether this model can give a reasonable phenomenon by analyzing the spectrum of the massive KK modes and their couplings to the matter confined at $y=0$.
With the Neumann boundary condition: $\partial_{z}h_{\mu\nu}|_{z=0}=0$, we can obtain the solutions for the Schr\"{o}dinger-like equation (\ref{Shrodinger-like eq}):
\begin{eqnarray}
\Psi_{n}(z) =\frac{1}{N_{n}}
   \left(z+\frac{1}{\beta}\right)^{\frac{1}{2}}
   \left[J_{0}\Big(m_{n}(z+\frac{1}{\beta})\Big)\right. \nonumber \\
       \left. +\alpha Y_{0}\Big(m_{n}(z+\frac{1}{\beta})\Big)
   \right], \label{Psinz1}
\end{eqnarray}
where $\frac{1}{\beta} \equiv \frac{1}{k}-\frac{2 M^{3}_{*}c_{2}}{\sigma_{-}}$, $\alpha=-\frac{J_{1}({m_{n}}/{\beta})}{Y_{1}({m_{n}}/{\beta})}$, and $N_{n}$ is the normalization constant.
The spectrum $m_{n}$ of the graviton KK modes can be obtained by the Neumann boundary condition at $z_{b}$: $\partial_{z}h_{\mu\nu}|_{z=z_{b}}=0$, which reads as
\begin{eqnarray}
\frac{J_{1}\left(m_{n}(z_{b}+{1}/{\beta})\right)}
     {J_{1}(m_{n}/\beta)}
 =\frac{Y_{1}\left(m_{n}(z_{b}+{1}/{\beta})\right)}
       {Y_{1}(m_{n}/\beta)}.
\end{eqnarray}

Since $k z_{b}\sim 10^{16}$ and $c_{2}\sim 1$, the first few KK modes satisfy the condition ${m_{n}}/{\beta}\ll 1$ and $\alpha\approx \frac{\pi}{8}(\frac{m_{n}}{\beta})^{2}\rightarrow 0$. Then the solution (\ref{Psinz1}) becomes
\begin{eqnarray}
\Psi_{n}(z)\approx\frac{1}{N_{n}}\left(z+\frac{1}{\beta}\right)^{\frac{1}{2}}
   J_{0}\left(m_{n}(z+\frac{1}{\beta})\right),
\end{eqnarray}
and the spectrum reads
$m_{n} \approx \frac{x_{n}}{z_{b}}$,
where $x_n$ satisfies $J_{1}(x_{n})=0$, and $x_{1}=3.83$, $x_{2}=7.02$, $\cdots$.
The normalization constant $N_{n}$ can be determined as follows:
\begin{eqnarray}
 1   =   \int^{z_{b}}_{-z_{b}} \Psi_{n}^2 dz
  \approx \frac{1}{N^{2}_{n}}z^{2}_{b}[J_{0}(m_{n}z_{b})]^2.
\end{eqnarray}
Using the approximation $J_{0}(x)\approx \sqrt{\frac{2}{\pi x}}\cos(x-\frac{1}{4}\pi)$ and the approximate formula of the zero point of $J_{1}(x_{n})$, $x_{n}\approx (n+\frac{1}{4})\pi$, we have
$N_{n}\approx\sqrt{\frac{2 z^{2}_{b}}{\pi x_{n}}}$.
Then the normalized KK modes are
\begin{eqnarray}
\Psi_{n}(z)\approx\sqrt{\frac{\pi x_{n}}{2 z_{b}^{2}}}
  \left(z+\frac{1}{\beta}\right)^{\frac{1}{2}}
  J_{0}\left(m_{n}(z+\frac{1}{\beta})\right).
\end{eqnarray}
The interaction between these KK modes and the matter confined on the positive tension brane is
\begin{eqnarray}
L^{(n)}_{\text{int}}={M^{-\frac{3}{2}}_{*}}\widetilde{T}^{\mu\nu}(x)h^{(n)}_{\mu\nu}(x,0)
 =\widetilde{\xi}^{(n)}\widetilde{T}^{\mu\nu}(x)\varepsilon^{(n)}_{\mu\nu}(x),
\end{eqnarray}
where $\widetilde{T}_{\mu\nu}$ is the symmetric conserved Minkowski space energy-momentum tensor. Then the couplings are
\begin{eqnarray}
\widetilde{\xi}^{(n)} \sim {M^{-\frac{3}{2}}_{*}}\sqrt{\frac{x_{n}}{\beta z^2_{b}}}
\sim \frac{\sqrt{n}}{10^{16}{\text{TeV}}} \sim \sqrt{n}~M^{-1}_{\text{Pl}}. \label{coupling KK}
\end{eqnarray}
The above result shows that the couplings of the first few KK modes to the matter have the similar strength to the coupling of the massless graviton to the matter.
{The mass spectrum of the KK modes is}
\begin{eqnarray}
 m_{n}\sim \frac{n}{z_{b}} \sim n~10^{-4}\text{eV}
\end{eqnarray}

To estimate experimental effects, we first consider the process that involves emission of gravitons, which could be observed as missing energy \cite{ArkaniHamed:1998rs, Antoniadis:1998ig, Rubakov:2001kp}.
The total cross section for this process with the center of mass energy $E_{\text{CM}}$ is of order
\begin{eqnarray}
 \sigma&\sim& \sum_{n} \left(\widetilde{\xi}^{(n)}\right)^{2}
 \sim \frac{n^{2}}{M^{2}_{\text{Pl}}}\sim \frac{(E_{\text{CM}}/10^{-4}\text{eV})^2}{M^{2}_{\text{Pl}}}\nonumber\\
 &\sim& \frac{E_{\text{CM}}^{4}}{(1 \text{TeV})^{4}}\frac{1}{E_{\text{CM}}^{2}}, \label{real emission}
\end{eqnarray}
which would have a significant increase when the energy approach 1TeV.
On the other hand, the massive KK modes with the mass gap $10^{-4}$eV would contribute an obvious deviation to Newtonian potential when distance less than the critical distance $R_c\sim 2$mm.
These results are the same as the six-dimensional ADD model's.

To understand the coincidence, we first note that in our five-dimensional warped model
the conformal size of the fifth dimension satisfies
\begin{eqnarray}
 M^2_{\text{Pl}}\sim M^{3}_{*}k z^{2}_{b}\sim M^{4}_{*}z^{2}_{b}
\end{eqnarray}
and in the six-dimensional ADD model the physical size $R$ satisfies
\begin{eqnarray}
M^2_{\text{Pl}}\sim M^{4}_{*} R^{2},
\end{eqnarray}
which leads to $z_{b}\sim R$.

\begin{figure}[htb]
\includegraphics[scale=0.7]{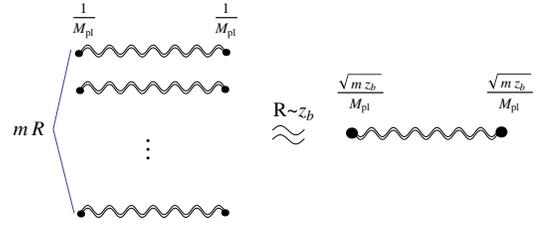}
\caption{The connection between propagators of KK gravitons in the six-dimensional ADD model (left) and the new five-dimensional warped model (right). \label{picture 2} }
\end{figure}

Further, in the six-dimensional ADD model, the degeneracy of KK modes with same mass $m$ is about
$mR$ and they have the same couplings of $M^{-1}_{\text{Pl}}$ to matter.
In our five-dimensional warped model, there is only one KK mode with mass $m$ due to only one extra dimension and the coupling to matter is $(m z_{b})^{\frac{1}{2}}M_{\text{Pl}}^{-1}$.
These two models have the same contribution to a process since
\begin{eqnarray}
m R(M_{\text{Pl}}^{-1})^2 \sim \left((m z_{b})^{\frac{1}{2}}M_{\text{Pl}}^{-1}\right)^{2}.
\end{eqnarray}

As illustrated in figure \ref{picture 2}, the total contribution of all the propagators of the KK modes in the six-dimensional ADD model is the same as that of the only propagator in our five-dimensional model.
So we can use the six-dimensional ADD model as a bridge to understand some phenomena in this new mechanism. However, phenomena involve the interactions among the graviton KK modes or related to scalar and vector modes may be different.

\section{General realization of the new mechanism in modified gravity}

The fundamental scale is also assumed as $M_*\sim 1$TeV and the standard model particles are confined on the brane at $y=0$.
A possible class of actions in five dimensions is
\begin{eqnarray}
S= \frac{M^{3}_{*}}{2} \int d^{5}x \sqrt{-g}
  \Big[  \psi \mathcal{R} + \mathcal{L}(g_{AB},\psi,\Phi_i)\Big]
          + S_{\texttt{b}},\label{dual action}
\end{eqnarray}
where only the field $\psi$ nonminimally couples to the curvature scalar, and $\Phi_i$ in the Lagrange density $\mathcal{L}$ denote other dynamic and/or nondynamic fields \footnote{For example see Ref.~\cite{Yang:2012dd}.}.
Assume that the system supports a family of solutions:
\begin{eqnarray}
a(y)=e^{-k|y|},~~\psi(y){\propto} e^{(N+2) k |y|},
\end{eqnarray}
where $k\sim M_*$ and now $N$ is just a dimensionless parameter determined by the dynamic of the theory.

Since only the field $\phi$ nonminimally couples to curvature scalar, the graviton zero mode can be calculated as \begin{eqnarray}
\Psi_0{\propto} a^{3/2}\psi^{1/2} {\propto} e^{\frac{1}{2}(N-1)k|y|}=(1+k|z|)^{(N-1)/2}
\end{eqnarray}
which will be diluted exponentially by the physical size of the extra dimension when $N>1$.
So in this case, the effective four-dimensional Planck mass will be warped up and be determined by
\begin{eqnarray}
M^{2}_{\text{Pl}}\sim M^{3}_{*}(kz_{b})^{N-1}z_{b}\sim M^{N+2}_{*}z_{b}^{N}.
\end{eqnarray}
This is exactly the same as in the $(4+N)$-dimensional ADD model, in which $M^{2}_{\text{Pl}}\sim M^{N+2}_{*} R^{N}$, if $z_{b}\sim R$.
It means that we can set the conformal size $z_{b}\sim R$ to solve the gauge hierarchy problem, while at the same time to keep a small physical length $y_b\sim \frac{74}{N}M_*^{-1}$ to avoid the remained hierarchy problem in the ADD model.

With the similar procedures as in previous section, we can solve the graviton KK modes
\begin{eqnarray}
\Psi_{n}(z)\approx\frac{1}{N_{n}}\left(z+\frac{1}{k}\right)^{\frac{1}{2}}J_{\frac{N}{2}-1}
\left(m_{n}(z+\frac{1}{k})\right).
\end{eqnarray}
The spectrum is given by $m_{n}\approx\frac{x_{n}}{z_{b}}$, where $x_{n}$ are determined by $J_{\frac{N}{2}}(x_{n})=0$. This means that the new mechanism shares the same mass spacing as the ADD model. The normalization constants are $N_{n}\sim {z_b}/\sqrt{{x_{n}}}$
and then
\begin{eqnarray}
\Psi_{n}(0)\sim {x^{\frac{N-1}{2}}_{n}}
                   {k^{\frac{1-N}{2}}z_{b}^{-\frac{N}{2}}}.
\end{eqnarray}
So the couplings of these KK modes to the matter confined at $y=0$ are given by
\begin{eqnarray}
\widetilde{\xi}^{(n)}\sim {M^{-\frac{3}{2}}_{*}}\Psi_{n}(0)
\sim {M_{\text{Pl}}^{-1}}   n^{\frac{N-1}{2}}.
\end{eqnarray}

In the $(4+N)$-dimensional ADD model, the { degeneracy} of the graviton KK modes with mass $m$ is about $(mR)^{N-1}$ and each mode has a coupling ${M_{\text{Pl}}^{-1}}$ to matter. In the new mechanism, there is only one graviton KK mode with mass $m$ and its coupling to matter is about ${M_{\text{Pl}}^{-1}}
{(m z_{b})^{\frac{N-1}{2}}}$.
Similar to the previous discussion, these two models have the same contribution to a process since
\begin{eqnarray}
(mR)^{N-1}({M_{\text{Pl}}^{-1}})^{2}=({M_{\text{Pl}}^{-1}}{(m z_{b})^{\frac{N-1}{2}}})^{2}.
\end{eqnarray}
So we can use the $(4+N)$-dimensional ADD model as a bridge to understand some phenomena of this new mechanism.
{ For example, the correction to the Newtonian potential $r^{-1}$ is of the form $r^{-(1+N)}$.}

Interestingly, the relation between the new mechanism and the $(4+N)$-dimensional ADD model tells us that the effects of multi flat extra dimensions may originate from the warped geometry with only one extra dimension.
Intuitively speaking, the massless graviton in warped geometry can be diluted as sparse as in multi flat extra dimensions.
One should note that in the new mechanism $N$ is a dynamically determined parameter and in the ADD model it is the number of extra dimensions. Because $N$ is dynamical in the new mechanism, different stages of the universe may have phenomena of different $N$.

The new mechanism also provides a new way to escape the experimental constrains.
Since the fundamental scale $M_{*}\sim k\sim 1$TeV, the critical distance of breaking the Newtonian inverse square law is given by
\begin{eqnarray}
R_{c}\sim 2\times 10^{\frac{32}{N}-16}\text{mm}.
\end{eqnarray}
Thus, to satisfy the experimental constrain $R_{c}\lesssim 0.1$mm~\cite{Adelberger:2003zx,Long:2002wn,Murata:2014nra}, we need
\begin{eqnarray}
N\gtrsim 2.18.
\end{eqnarray}
{Since in the ADD model, $N$ must be an integer and the least possible choice is $N=3$. However, $N=3$ leads to a
deviation of Newtonian inverse square law starts from $R_{c}\sim 10^{-5}$mm, which can't be observed easily.
Since $N$ is a dynamically determined parameter in our new mechanism, non-integer $N$ is allowed. So our new mechanism can satisfy the experimental constraints more easily and give interesting observable effects.
Since the correction to the Newtonian potential $r^{-1}$ is of the form $r^{-3.18}$ in the new mechanism and $r^{-3}$ in the six-dimensional ADD model, these two models can be distinguished in experiments.}
\section{Conclusion and discussion}
In order to avoid the remained hierarchy in the ADD model, there are two kinds of mechanisms by using the warped geometry.

One is the RS1 model and its generalization. In this kind of model, the massless graviton is localized near $y=0$ and will not be diluted exponentially by the size of the extra dimension, so the effective four-dimensional Planck scale is remain the same as the fundamental five-dimensional Planck scale. Thus, the standard model particles should be confined at $y=y_{b}$ to redshift the electroweak scale to solve the gauge hierarchy problem.

Another one is the new mechanism proposed in this { paper}. In this  mechanism, the massless graviton is localized near $y=y_{b}$ and is diluted exponentially, so the effective Planck scale grow exponentially with the size of the extra dimension. Thus, we can confine the standard model particles at $y=0$ to make the electroweak scale remain the fundamental scale.
It is a mechanism with the size of the extra dimension a few TeV$^{-1}$, different from the $M_{\text{Pl}}^{-1}$ in the RS1 model.

{In the explicit realization in the nonlocal gravity theory, the tower of spin-2 excitations has mass gap $10^{-4}\text{eV}$ and these KK gravitons are coupled with the gravitational scale to the standard model particles, while in the RS model, both the mass gap and the coupling are $\text{TeV}$ scale. We further discussed the possible generalizations in other modified gravity theories. We found that the new mechanism would lead to reasonable phenomena similar to the $(4+N)$-dimensional ADD model. This implies that the phenomena of flat extra dimensions could be emerged from warped geometry with only one extra dimension. Besides, it provides a new way to build braneworld model that satisfies the experimental constraints since $N$ can be a non-integer.}

This new mechanism with warped extra dimension was constructed in nonlocal gravity and a general class of scalar tensor gravities. Since a large class of modified gravities has an extra scalar degree of freedom, we believe that this kind of construction widely exists in modified gravities.

\section*{Acknowledgement}

This work was supported by the National Natural Science Foundation of China (Grants No. 11875175, No. 11747021, No. 11675064, and No 11522541).

\end{document}